# Observation of the Breakdown of Optical Phonon Splitting in a Two-dimensional Polar Monolayer

Jiade Li[1,2], Li Wang[1], Zhiyu Tao[1,2], Weiliang Zhong[1,2], Siwei Xue[1], Guangyao Miao[1], Weihua Wang[1], Jiandong Guo[1,2*], & Xuetao Zhu[1,2*]

[1] *Beijing National Laboratory for Condensed Matter Physics and Institute of Physics, Chinese Academy of Sciences, Beijing 100190, China*

[2] *School of Physical Sciences, University of Chinese Academy of Sciences, Beijing 100049, China*

[*] Jiandong Guo ( jdguo@iphy.ac.cn ) and Xuetao Zhu ( xtzhu@iphy.ac.cn ).

**Phonon splitting of the longitudinal optical and transverse optical modes (LO-TO splitting), a ubiquitous phenomenon in three-dimensional (3D) polar materials, is essential for the formation of the 3D phonon polaritons. Theories predict that the LO-TO splitting will break down in two-dimensional (2D) polar systems, but direct experimental verification is still missing. Here, using monolayer hexagonal boron nitride (*h*-BN) as a prototypical example, we report the direct observation of the breakdown of LO-TO splitting and the finite slope of the LO phonons at the center of the Brillouin zone in 2D polar materials by inelastic electron scattering spectroscopy. Interestingly, the slope of the LO phonon in our measurements is lower than the theoretically predicted value for a freestanding monolayer due to the screening of the Cu foil substrate. This enables the phonon polaritons (PhPs) in monolayer *h*-BN/Cu foil to exhibit ultra-slow group velocity ($\sim 5 \times 10^{-6} c$, $c$ is the speed of light) and ultra-high confinement (~ 4000 times smaller wavelength than that of light). Our work reveals the universal law of the LO phonons in 2D polar materials and lays a physical foundation for future research on 2D PhPs.**

Optical phonons, the out-of-phase collective vibrations of lattice, play essential roles in the optical[1,2], electronic[3,4], and thermal[5,6] properties of crystalline materials. In particular, the behaviors of longitudinal optical (LO) phonons are distinctive due to the polarity of the materials. In non-polar materials, whether in three-dimensional (3D) or two-dimensional (2D) systems, the lattice symmetry guarantees that the LO and the transverse optical (TO) phonons are degenerate with zero slopes at the center of the Brillouin zone (CBZ) (Fig. 1a). However, in polar materials, the lattice vibrations of the LO phonons generate extra long-range electric fields, which in turn exert long-range Coulomb forces on the polar lattices. The long-range Coulomb interaction significantly changes the behaviors of the LO phonons. In the 3D scenario, the long-range Coulomb interaction raises the frequency of the LO phonon near the CBZ, leading to an energetic split with the TO phonon, known as the LO-TO splitting (Fig. 1b). The LO-TO splitting ubiquitously exists in 3D polar materials, with experimental observations reported in GaP[7], SiC[8], BN[9], *etc*. In the 2D scenario, however, the LO phonon does not split with the TO phonon anymore. In the last two decades, various theoretical models predict that the LO phonon degenerates with the TO phonon in 2D monolayers and exhibits a finite slope at the CBZ (Fig. 1c)[10-14]. It is worth noting that the LO-TO splitting is essential for the formation of the 3D phonon polaritons (PhPs)[15]. The absence of the LO-TO splitting in 2D polar materials raises a fundamental question about the nature of the 2D PhPs in polar monolayers[16].

The 2D PhPs are predicted to have excellent characteristics such as ultra-slow group velocity and ultra-high wavelength confinement of light, and thus can serve as an ideal platform for strong light-matter interaction[16,17]. However, measuring the dispersion of the 2D PhPs in a real monolayer with current mainstream techniques, such as the scanning near-field optical microscopy (s-SNOM)[18] and the electron energy loss spectroscopy incorporated in a scanning transmission electron microscope (STEM-EELS)[19], is still challenging due to their small momentum compensation. Although a weak 2D PhP signal has been detected in recent s-SNOM measurements[20], the momentum window is too narrow to reveal the overall properties of the 2D PhPs. Fortunately, recent theoretical studies have connected the microscopic phonon properties with the macroscopic electromagnetic response, demonstrating that the 2D PhPs are simply equivalent to the LO phonons in 2D polar materials[16]. In this regard, a measurement of LO phonons in 2D polar materials provides a new methodology for studying the optical properties of the 2D PhPs. Therefore, the direct experimental observation of the behaviors of the LO phonons in strictly 2D polar monolayers is urgently needed, whether to verify the breakdown of the LO-TO splitting or to study the properties of the PhPs in 2D.

Monolayer hexagonal boron nitride (*h*-BN), a prototypical 2D polar material, is an ideal candidate for the experimental study of the behaviors of the LO phonons and PhPs in 2D monolayers. Especially, a recent theory predicts[17] that the 2D PhP of a freestanding monolayer *h*-BN exhibits an ultra-slow group velocity (~ $10^{-5}$ *c*, *c* is the speed of light) and ultra-high confinement (~ 1000 times smaller wavelength than that of light in free space), indicating great potential for optoelectronic applications. High-resolution electron energy loss spectroscopy (HREELS), highly sensitive to excitations on surfaces[21], is an ideal method for measuring phonon dispersions of 2D monolayers. Especially, the development of 2D-HREELS[22,23] that can perform 2D measurements of energy and momentum simultaneously with high resolutions (Methods), brings new opportunities to observe the behaviors of the LO phonons in 2D polar materials. Here, taking monolayer $h$-BN as a prototypical example, using the 2D-HREELS measurements combined with theoretical calculations, we reveal the dispersion behaviors of the LO phonons and the properties of the PhPs in 2D polar monolayers.

**Sample characterization and phonon spectra**

Monolayer $h$-BN was synthesized on a substrate of polycrystalline copper foil ($h$-BN/Cu foil) (Methods). As shown in Fig. 2a and 2b, the high crystallographic quality of $h$-BN was confirmed by the atom-resolved scanning tunneling microscope (STM) image and sharp low-energy electron diffraction (LEED) pattern. The 2D-HREELS measurements were performed at room temperature using an incident electron beam with an energy of 110 eV and an incident angle of 60°. Figure 2c shows the 2D energy and momentum mappings for the phonon spectra of the monolayer $h$-BN along the Γ−M and Γ−K directions. Unlike the HREELS spectra measured for graphene[24,25] where the non-polar nature makes the phonon intensity distribution relatively uniform in the momentum space, the signal intensity of the phonon dispersions of $h$-BN in the impact scattering region (away from the Γ point, Fig. 2f and 2g) is suppressed by the strong dipole scattering (around the Γ point, Fig. 2e) due to the polarity of $h$-BN. Nevertheless, we managed to capture all the phonon modes of monolayer $h$-BN (see Fig. 2d, the corresponding second derivative results of Fig. 2c). To assign the phonon modes measured by 2D-HREELS, the calculated phonon dispersions, with the 2D implementation of the nonanalytical-term correction[26,27] (green curves, see Methods for details), are also superimposed in Fig. 2c and 2d. The corresponding LO, TO, out-of-plane optical (ZO), transverse acoustic (TA), longitudinal acoustic (LA), and out-of-plane acoustic (ZA) modes are labeled in the calculated phonon dispersions. It can be seen from Fig. 2d that the experimental results are in good agreement with the calculations, and all six phonon modes are identified along the Γ−K direction (although the signal intensity of the TA mode is very weak). Especially, our calculated LO phonon of monolayer $h$-BN exhibits a "V-shaped" nonanalytical behavior near the Γ point, which originates from long-range Coulomb interactions in 2D polar materials, reproducing the results of previous studies[12,28].

We also notice that there are some dispersionless scattering signals in Fig. 2d (marked by the purple dashed line), which show obvious loss peaks in the energy distribution curves (EDCs) (marked as β peaks in Fig. 2f and 2g). After careful analyses (discussed in detail in the Supplementary Information), these signals are proven to be the phononic replicas caused by the surface roughening of the Cu foil substrate rather than the true phonon dispersions of monolayer $h$-BN. Due to the limitation of the commercial cold rolling process and the flexibility of thin Cu foil, the surface roughness of the substrate is unavoidable. Fortunately, the phononic replicas from the substrate roughness are uniformly distributed in the momentum space like diffuse scattering[21], and thus the dispersions of the monolayer *h*-BN are unaffected and still well discernable.

**Breakdown of the LO-TO splitting**

To illustrate the measured dispersion of the LO phonon of monolayer *h*-BN, in Fig. 3a we show a zoom-in view of Fig. 2d around the LO phonon. The dispersion of the LO phonon is unambiguously demonstrated and shows a distinctly exotic behavior near the Γ point. First, the LO and TO phonons degenerate undoubtedly at the Γ point (see also in Fig. 2e for the EDC at the Γ point). These two branches gradually separate from each other as the momentum increases away from the Γ point. Second, the dispersion of the LO phonon is isotropic along the Γ–M and Γ–K directions, and shows a finite positive slope at the Γ point. The dispersion of the LO phonon does exhibits a "V-shaped" nonanalytic behavior near the Γ point, which agrees well with the calculations. Our observation, with experimental visualization of the phonon dispersions, directly verifies the physical picture shown in Fig. 1c, and thus proves that the LO-TO splitting does breakdown in 2D monolayer *h*-BN.

Furthermore, we compare the calculated phonon spectra of monolayer *h*-BN with or without considering the long-range Coulomb interaction. Notably, the calculated results considering the long-range Coulomb interaction (green curves in Fig. 3a) show good agreement with the measurements, while the results without considering the long-range Coulomb interaction (gray curves in Fig. 3a) have obvious deviations from the measurements. This demonstrates that the long-range Coulomb interaction clearly affects the dispersion of the LO phonons in 2D polar materials. Note that when the momentum $q > 0.5$ Å$^{-1}$, both the two calculated results show the same trend as the experimental results, suggesting that the modulation of the long-range Coulomb interaction becomes insignificant at large momenta. Our experimental results demonstrate the correctness of the 2D implementation of the nonanalytical-term correction[12,27] and lay the foundation for a correct understanding of the behavior of the LO phonons in 2D polar monolayers.

**Screening effect from Cu foil**

The behaviors of the LO phonons in the polar monolayers can be strongly affected by electronic screening. Figure 3b shows the phonon dispersion (red balls) extracted from the 2D-HREELS measurement. Interestingly, our results show that the slope of the LO phonon at the Γ point is about $5 \times 10^{-6}\ c$, which is much lower than the theoretical prediction of $1.2 \times 10^{-4}\ c$ for freestanding monolayer *h*-BN[16]. Theory predicts that the screening behavior in the 2D case is very different from the 3D due to the dimensionality dependence of the long-range Coulomb interaction produced by the macroscopic polarization[14,29,30]. On the other hand, the application of 2D polar materials is always based on the support of the substrates. So, it is very important to study the screening effect from the substrate on the 2D polar materials. In our study, the monolayer $h$-BN is grown on the Cu foil substrate, thus providing a natural platform to study the screening from the substrate on the LO phonons. To evaluate the screening effect, we fit the experimental dispersion using the following model[12]

$$\omega_{LO}(q) = \sqrt{\omega_{TO}^2(q=0) + \frac{S\,q}{\varepsilon_{env} + r_{eff}\,q}} \qquad (1)$$

with

$$\varepsilon_{env} = \frac{\varepsilon_{top} + \varepsilon_{bot}}{2} \qquad (2)$$

where $\omega_{LO}(q)$ is the LO phonon frequency at momentum $q$, $\omega_{TO}(q=0)$ is the TO phonon frequencies at $q=0$, $S$ is a parameter related to the Born effective charges and phonon displacements, $r_{eff}$ is the effective screening length describing the screening properties of the 2D material itself, $\varepsilon_{env}$ is the average dielectric constant of the surrounding environment, and $\varepsilon_{top}$ ($\varepsilon_{bot}$) is the effective dielectric constants of the top (bottom) side of the 2D polar materials. From our measurement results, we have $\omega_{TO}(q=0) = 172.5$ meV. $S$ and $r_{eff}$ are material-specific properties independent of the surrounding environment. Here, we set $S = 8.40 \times 10^{-2}$ eV$^2 \cdot$ Å and $r_{eff} = 7.64$ Å for monolayer $h$-BN obtained from the calculations with density functional theory[12]. In the current system, the top side of the monolayer $h$-BN is vacuum, and thus we have $\varepsilon_{top} = \varepsilon_{vacuum} = 1$. Therefore, we have only one independent variable $\varepsilon_{bot}$ contributed by the Cu foil on the bottom of the $h$-BN. Figure 3b shows the fitting results (origin curves). The effective bottom dielectric constant extracted from the fitting is $\varepsilon_{bot} = 4.45$.

In our study, the effective dielectric constant originates from the nonlocal effects of the Cu foil induced by the LO phonon of the monolayer $h$-BN. Nonlocal screening from the metal substrate can significantly change the plasmonic or phononic response of the materials[31,32]. For comparison, the calculated dispersion curves for $\varepsilon_{bot} = 1$ and $10$ are also plotted in Fig. 3b, indicating that the dispersion of the LO phonon is significantly modulated by the screening of environment dielectric properties. Compared to the freestanding monolayer $h$-BN ($\varepsilon_{bot} = 1$), the effect of the nonlocal screening of the Cu foil substrate significantly reduces the slope of the LO phonon.

**Properties of 2D PhPs**

The observed breakdown of the LO-TO splitting fulfills the prerequisite for the existence of 2D PhPs in polar monolayers. As theoretically demonstrated, the LO phonons are simply equivalent to the 2D PhPs in polar monolayers under the premise of the breakdown of the LO-TO splitting[16]. Thus, information on the 2D PhP of monolayer $h$-BN can be directly derived from our measurements of the LO phonon dispersion. Two key figures of merit about the 2D PhPs in $h$-BN we investigate here are the deceleration factor and the confinement factor. The deceleration factor is a measure for the slowdown of light trapped in a PhP and is defined by the ratio of the group velocity of the PhP to the speed of free light, $v_g = c^{-1} \cdot \partial\omega/\partial q$, where $\omega$ is the frequency of the PhP at a specific momentum $q$. The confinement factor is a measure for the compression of the wavelength of light trapped in a PhP and is defined by the ratio of the momentum of PhP to the wavevector of free light, $q/q_0$, where $q_0$ is the wavevector of free light at the corresponding PhP frequency. Figures 4a and 4b show the comparison of the deceleration factors and confinement factors of $h$-BN for thick-layer (10 nm, data extracted from ref.[33]), freestanding monolayer [calculated by Eq. (1)], and monolayer on Cu foil (our results). The PhP in monolayer $h$-BN exhibits lower deceleration and larger confinement factors than the thick-layer. This can be easily deduced from the dependence of the dispersion of $h$-BN PhPs on the number of layers (Fig. 4 c). Under a specific energy, thicker layers always have higher dispersion slope and smaller momentum than thinner layers, which gives the thinner layers a lower deceleration factor and higher confinement factor. In this regard, the PhP in monolayer $h$-BN has the optimal deceleration factor and confinement factor. Shockingly, our experimental results show that the PhP of monolayer $h$-BN on Cu foil has lower group velocity and higher confinement factor than the calculated results of the freestanding monolayer $h$-BN. The PhP of $h$-BN/Cu foil exhibits an ultra-slow group velocity down to about $5 \times 10^{-6}\ c$ at the long-wave limit and an ultra-high confinement factor up to about 4000, surpassing the existing reported PhP records of the lowest group velocity (~ $10^{-5}\ c$) and the highest confinement factor (~ 500) (ref.[33]), respectively.

The excellent properties of the 2D PhP of monolayer $h$-BN in our study benefit from the nonlocal screening of Cu foil substrate. Although tuning the PhPs of thick-layer $h$-BN via the screening of metallic substrate has been extensively investigated[34-37], it has not been explored experimentally in strictly monolayer $h$-BN. Our study shows that the Cu foil provides a suitable screening strength, which reduces the group velocity of the 2D PhP while retaining the polarized electric fields of monolayer $h$-BN. This enables the monolayer $h$-BN/Cu foil system to exhibit ultra-slow group velocity and ultra-high confinement of light beyond the 2D limit.

**Comparison of the methods for measuring PhPs**

Currently, the mainstream methods for measuring PhPs are s-SNOM and STEM-EELS, both of which obtain the dispersion information of PhPs through real-space measurements. The PhPs are excited by an incident beam (light or electron) near the sample boundary. The excited PhPs propagate to the edge, and are reflected by the edge. And then the reflected and the original PhPs interfere with each other, leading to a maximum interference intensity at

$$2qd + \varphi = 2\pi \qquad (3)$$

where $d$ is the distance to the edge boundary, $\varphi$ is the phase change introduced by the reflection[38,39]. Here we set $\varphi = \pi/4$ according to ref.[39]. The dispersion of a PhP can be obtained by measuring the excitation energy and $d$ according to Eq. (3). We need to note that in Eq. (3), $q$ is inversely proportional to $d$. This makes the PhPs with very sharp dispersions instead exhibit very flat spatial distributions, and vice versa (see the dispersions and spatial distributions of PhPs for 100 layers and monolayer $h$-BN in Fig. 4c and the inset of Fig. 4d). In this sense, the tools that perform measurements in real-space and momentum-space are complementary in measured precision and range. In thick-layer samples, the sharp dispersion of PhPs at small momentum makes it challenging for 2D-HREELS measurements, but its flat spatial distribution enables s-SNOM/SETM-EELS to measure the dispersion of PhPs with high precision[18,19,33,38-40]. In contrast, in monolayer samples, the group velocity of the LO phonon is small and can be easily captured by 2D-HREELS, but the spatial distribution of 2D PhP is very sharp. As shown in Fig. 4d, converting our measured results to the spatial distribution of PhP near the $h$-BN sample boundary by Eq. (3), we can see that the sharp change of PhP energy with $d$ occurs within ~ 6 nm. Considering the strong localization and defects at the sample boundary[33] and the spatial resolution of the s-SNOM/STEM-EELS, it is extremely challenging to perform spatially resolved measurements of 2D PhPs within such small $d$. In this regard, 2D-HREELS provides a new methodology for investigating the properties of 2D PhPs.

It is worth noting that STEM-EELS has also demonstrated the ability to measure the phonon dispersion of thick-layer $h$-BN[41,42], but it is insensitive to the monolayers and has lower energy resolution, making it difficult to capture phonon dispersion in strictly monolayer $h$-BN. In addition, we also noticed that there have been reports on the measurement of monolayer $h$-BN phonon dispersion by the conventional HREELS[43], but it could not distinguish the dispersion of LO and TO branches near the Γ point due to insufficient momentum resolution. Our results demonstrate that the 2D-HREELS, satisfying both surface sensitivity and high energy and momentum resolutions, is a unique method for measuring phonon behaviors and PhP properties in polar monolayers.

**Conclusion**

By measuring the phonon spectra of monolayer $h$-BN with high energy and momentum resolutions, we observe the breakdown of LO-TO splitting at the CBZ. Combined with first-principles calculations, it is demonstrated that this exotic behavior originates from the long-range Coulomb interactions of lattice vibrations. We also find that the screening from the substrate can reduce the slope of the LO phonon, which makes the 2D PhP of $h$-BN show an ultra-slow deceleration factor (~ $5 \times 10^{-6}\ c$) and ultra-high confinement factor (~ 4000) beyond the 2D limit. This indicates that monolayer $h$-BN/Cu foil is an ideal system for exploring light-matter interactions. Last, we illustrate the complementarity of measuring PhPs in real-space and momentum-space, providing a new methodology for studying the physical properties of 2D PhPs using 2D-HREELS. By studying the extraordinary properties of 2D PhPs using this innovative methodology, we bring new opportunities for optoelectronic applications of 2D polar monolayers.

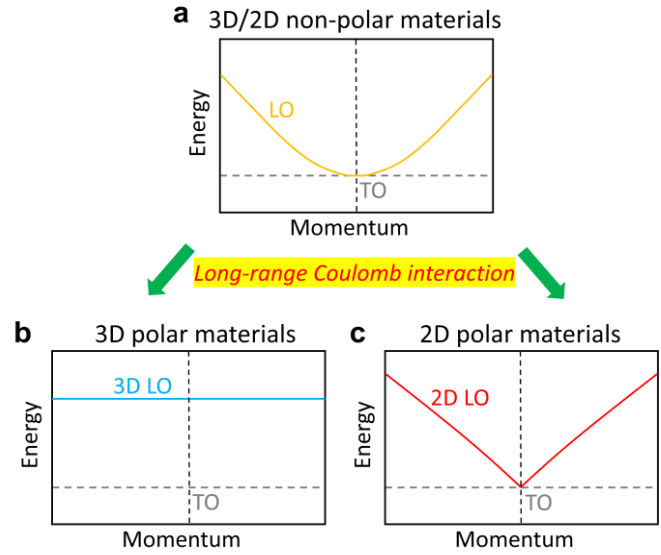

**Fig. 1| Schematic of the behaviors of the LO phonons near the CBZ. a-c**, The behaviors of the LO phonons in 3D/2D non-polar, 3D polar, and 2D polar materials, respectively. The dashed black lines represent the CBZ.

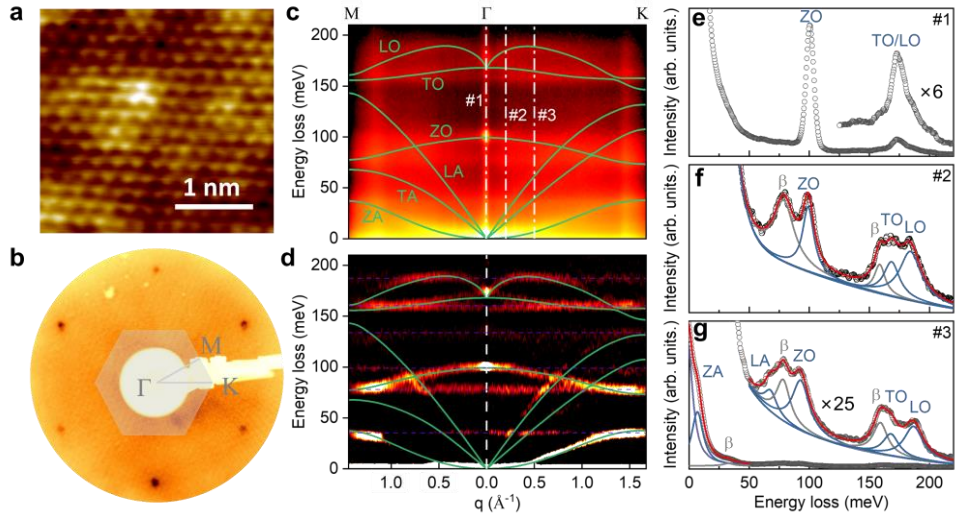

**Fig. 2| Crystallographic quality and phonon spectra of monolayer *h*-BN. a**, Atomic-resolution STM image (scanned at 0.9 V and 0.3 nA). **b**, LEED pattern, obtained at room temperature with an incident beam energy of 140 eV. **c**, 2D energy-momentum mappings of 2D-HREELS along the Γ−M and Γ−K directions. **d**, The second derivative results correspond to **b**. The purple dashed lines mark the replica signals of the phonon. **e**, **f**, **g**, EDCs corresponding to the #1 (specular direction), #2 (momentum 0.2 Å$^{-1}$), and #3 (momentum 0.5 Å$^{-1}$) dashed dot lines in **b**, respectively. The green curves in **c** and **d** are calculated phonon dispersions by considering the 2D implementation of the nonanalytical-term correction[26,27].

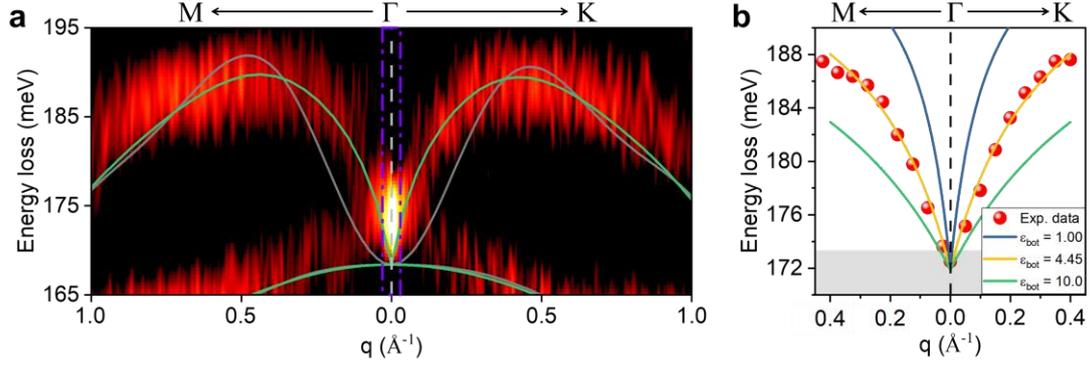

**Fig. 3| The dispersions of the LO phonon along the M–Γ–K direction. a**, Comparison between the measured 2D phonon spectra of the LO phonon and the calculated phonon dispersions. Green curves correspond to the calculated results in Fig. 2b. Gray curves are the calculated phonon dispersions without considering the nonanalytical-term correction. **b**, Comparison between the measured (red balls) dispersion of the LO phonon and the calculated results (solid curves) with $\varepsilon_{bot} = 1$, 4.45, and 10. $\varepsilon_{bot} = 1$ represents the freestanding $h$-BN, $\varepsilon_{bot} = 4.45$ is the result of the best fit to the experimental dispersion. The height of the gray shaded area represents the full width at half maximum of the zero-loss peak.

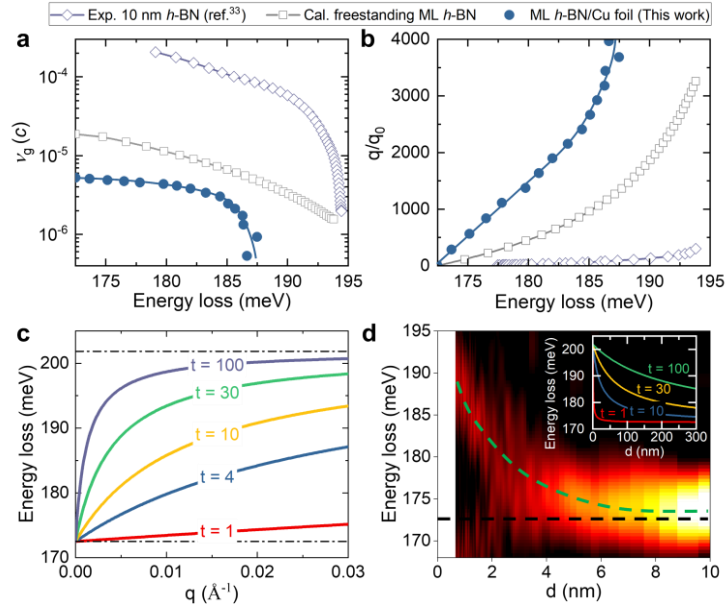

**Fig. 4| 2D PhPs of monolayer *h*-BN. a**, **b**, Deceleration and confinement factors versus energy loss, respectively. Solid circles represent the results of this work, empty squares represent the calculated results for a freestanding monolayer (ML) by Eq. (1), and empty rhombus represent the measured results of 10 nm-thick extracted from ref.[33]. **c**, The calculated dispersion relation of the PhPs for various freestanding layers. Calculations performed using $\omega_{LO}^{t}(q) = \sqrt{\omega_{TO}^{2}(q=0) + t\,S\,q/(1 + t\,r_{eff}\,q)}$ (ref.[12]), where $t$ is the number of layers. **d**, Spatial distribution of PhP transformed from 2D phonon spectra measured along Γ−K direction (0–0.4 Å$^{-1}$). Inset: calculated spatial distribution of PhPs for various layers using Eq. (3).

## Methods

### Sample preparation

A polycrystalline Cu foil substrate with a thickness of 25 μm (sourced from Sichuan Oriental Stars Trading Co. Ltd) was employed in the synthesis of monolayer $h$-BN film. Prior to growth, the Cu foil was subjected to a heat treatment in a tube furnace, where it was heated to 1,050 °C over a period of one hour and annealed for an additional hour in a mixed gas flow of Ar (500 sccm) and $H_2$ (50 sccm) at atmospheric pressure. The precursor of ammonia borane with 97% purity (sourced from Aldrich) was placed in a ceramic crucible situated 1 m upstream from the substrate in the heating zone of the tube furnace. The system was then switched to low pressure (approximately 200 Pa) and subjected to a mixed gas flow of Ar (10 sccm) and $H_2$ (40 sccm). The precursor was heated to 65 °C within a period of 10 minutes and held at this temperature for 1.5 hours using a belt heater, allowing for sublimation and the subsequent growth of $h$-BN film on the substrate. After growth, the system was cooled naturally to room temperature under a gas flow consisting of Ar (500 sccm) and $H_2$ (10 sccm) at atmospheric pressure.

After being transferred to the 2D-HREELS system, the sample was annealed at 400 °C to remove possible contamination.

### HREELS measurements

In a HREELS measurement, the energy ($E_{loss}$) and momentum ($q$) of the phonons are obtained by the conservation of energy and momentum for the incident and scattered electrons. As given by

$$q = \frac{\sqrt{2mE_i}}{\hbar}\sin\theta_i - \frac{\sqrt{2mE_s}}{\hbar}\sin\theta_s \quad (4)$$

and

$$E_{loss} = E_i - E_s \quad (5)$$

where $\theta_i$ ($\theta_s$) and $E_i$ ($E_s$) are the incident (scattered) angles and energies of electrons, respectively. The conventional HREELS collects the energy loss curves of scattered electrons at a fixed angle in each measurement and the dispersion relation is achieved by rotating the monochromator, analyzer, or sample. Our developed 2D-HREELS can directly obtain a 2D energy-momentum mapping simultaneously by a specially designed lens system with a double-cylindrical Ibach-type monochromator combined with a commercial VG Scienta hemispherical electron energy analyzer. The HREELS and LEED are equipped in an ultrahigh-vacuum system with a base pressure better than $2 \times 10^{-10}$ Torr. The measurement direction of the sample is precisely controlled by the electric motors and examined by the LEED.

### STM measurements

The STM experiments were performed after HREELS measurements. The sample was transferred *ex-situ* into the ultra-high vacuum STM system (Unisoku) with a base pressure better than $2 \times 10^{-10}$ Torr. After being transferred to the STM system, the sample was also annealed at 400 °C. The morphology characterization of $h$-BN/Cu was performed at 78 K. Pt/Ir tips were used in the STM experiments.

**First-principles calculation**

The phonon dispersion of monolayer $h$-BN was calculated using the DFPT[44] implemented in the Quantum ESPRESSO[45,46]. In particular, the 2D implementation of the nonanalytical-term correction[12,27] is performed to get the correct dispersion of the LO phonon. We use the Perdew-Burke-Ernzerhof (PBE) exchange-correlation functional[47] combined with the Projector Augmented Wave (PAW) pseudopotentials[48]. A 55 Ry kinetic energy cut-off, a 12 × 12 × 1 electron momentum grids, and an 8 × 8 × 1 phonon momentum grids are applied. The optimized lattice parameters of $h$-BN are 2.509 Å, which shows excellent consistency with the earlier works[49,50].

**Data availability**
The data that support the findings of this study are available from the corresponding authors upon reasonable request.

**Acknowledgments**
This work was supported by the National Key R&D Program of China (Grants No. 2021YFA1400200 and No. 2022YFA1403000), the National Natural Science Foundation of China (Grant No. 12274446), and the Strategic Priority Research Program of the Chinese Academy of Sciences (Grant No. XDB33000000).

**Author contributions**
J. L. and Z. T. performed the HREELS experiments. J. L. performed the theoretical calculation and analysis. L. W. grew the sample. W. W., W. Z., and G. M. performed the STM experiments. S. X. participated in the data analysis and discussion. J. L., Z. X., and J. G. wrote the manuscript with input from all authors. Z. X. and J. G. supervised the project. All authors contributed to the discussion of the results.

**Competing interests**
The authors declare no competing interests.

## Supplementary Information

**Origin of replica signals in HREELS spectra**

In this section, we will demonstrate that the replica signals originate from the roughness of the $h$-BN/Cu foil surface. The $h$-BN grown Cu foil is rough due to the surface roughening of Cu foil [Fig. S1(a)]. Considering the incident beam size of our HREELS is around 1 mm, the scattering geometry cannot be perfect on a rough surface. As shown in Fig. S1(b), according to Eq. (4)-(5) in Methods, if the sample surface is atomically flat, for a specific $\theta_i$, a specific $q$ occurs only at a specific $\theta_s$, that is, one-to-one correspondence between $q$ and $\theta_s$. In this case, the true momentum ($q_r$) of the phonon is equal to the momentum transfer ($q_m$) measured by HREELS. However, if the sample surface is not flat enough, the roughness will produce small facets with different orientations. Correspondingly, the angle between the incident direction and the facet normal, *i.e.*, the $\theta_i$, varies with the facet orientations. Therefore, for a specific $q_r$, $\theta_s$ also varies with the facet orientations [Fig. S1(c)]. This means that the phononic scattering signal at a specific $q_r$ contributed by many small facets will be replicated to various $q_m$. This produces the replica signals of phonons uniformly distributed in momentum space.

In general, the overall intensity of the replica signals contributed by the small facets is related to the roughness of the sample surface. The intensity distributions of the replica signals in the 2D HREELS spectra are affected by the following two factors:

1) Intrinsic phonon dispersion of materials. At the energy position where the slope of the phonon dispersion is very small, the phonon frequency is almost constant in a certain momentum range. This enables the replica signals in this momentum range to be superimposed, increasing the intensity of the replica signals.

2) Selection rules for HREELS. The selection rule for HREELS determines the scattering intensity of phonons. The phonons with greater scattering intensity will produce greater intensity of the replica signals.

In our measurements, the replica signals originating from small facets are all related to phonons with larger scattering intensity and smaller dispersion slope (Fig. S2), which is consistent with the above discussions. Furthermore, according to our calculations using Eq. (4), for a phonon mode at $q_r = 1 \text{ Å}^{-1}$, a facet with a roughness angle as small as 10° will make the replica signal cover the entire first Brillouin zone. Thus, the roughness of the Cu foil substrate can easily give rise to the replica signals throughout the 2D phonon spectra of $h$-BN measured by HREELS.

It should be emphasized that in our measurements of $h$-BN/Cu foil, there is a main scattering plane on the sample surface, which contributes most of the intensity of the scattered signal. Figure S3(a) shows the 2D mapping near the zero-loss energy along the Γ-K direction. There is a bright spot with the maximum scattering intensity around the position of zero energy loss and zero momentum in the 2D mapping, corresponding to the specular scattering of electrons from the main plane. To show the dependence of the scattering intensity on the momentum at zero loss energy, we extracted the momentum distribution curve (MDC) and displayed it in Fig. S3(b). The extremely strong intensity of zero loss peak along the momentum direction indicates that most of the specular scattering signal comes from one main plane (specular scattering from other small facets contributes to a finite-momentum background in MDC). Correspondingly, most of the inelastic scattering signals in our measurements are also contributed by the main plane, which reflects the true phonon dispersions of $h$-BN. This is also demonstrated by the good agreement between our measured and calculated phonon dispersions. Therefore, the small scattering facets of the sample can only bring some replica signals and will not affect the real phonon dispersions from the scattering of the main plane.

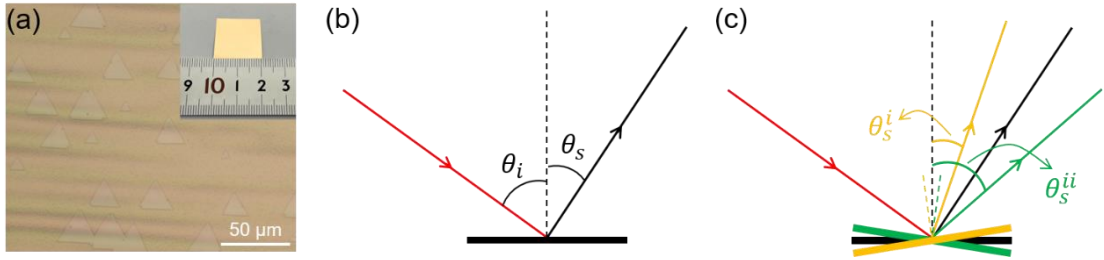

**FIG. S1. Sample characteristic and HREELS scattering geometry.** (a) Optical image of sub-monolayer *h*-BN grown on Cu foil. The image shows apparent roughness caused by the cold rolling process. (b) Schematic of the HREELS scattering geometry on a flat plane. (c) Schematic of the HREELS scattering geometry for a specific $q_r$ on different small facets.

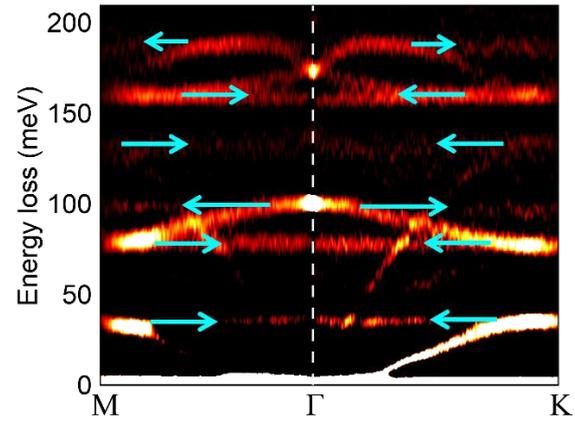

**FIG. S2. Origin of replica signal in HREELS spectra.** The bright blue arrows mark the phonon modes corresponding to the replica signals.

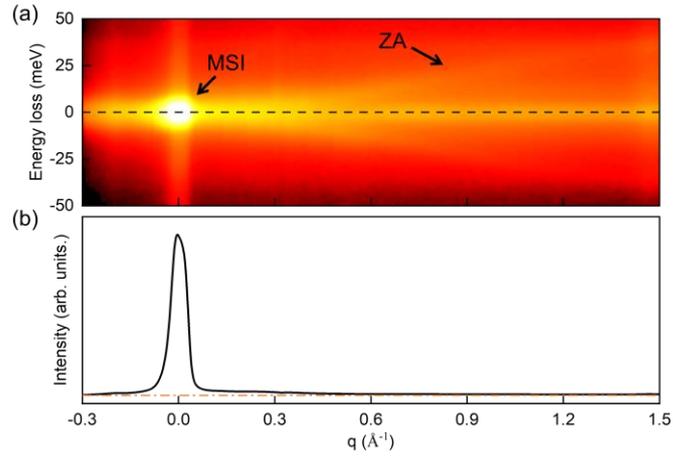

**FIG. S3. Specular scattering at zero-loss energy of the monolayer *h*-BN/Cu foil.** (a) 2D mapping near the zero-loss energy along Γ-K direction. The position with the maximum scattering intensity is marked with MSI. (b) MDC at the zero-loss energy [corresponding to the black dotted line in (a)].